\documentclass[twocolumn,superscriptaddress,groupedreferences]{revtex4-2}  
\usepackage{graphicx, dcolumn, bm, amssymb, epstopdf, soul, amsmath}
\usepackage[normalem]{ulem}
\usepackage{color}
\usepackage{xurl}

\newcommand\redsout{\bgroup\markoverwith{\textcolor{blue}{\rule[0.5ex]{2pt}{0.4pt}}}\ULon}

\usepackage{amsmath}
\usepackage{mathrsfs}  

\begin{document}
\widetext

\title{Mass testing and proactiveness affect epidemic spreading}
\author{Saptarshi Sinha} \author{Deep Nath} \author{Soumen Roy} 
\email{soumen@jcbose.ac.in}
\affiliation{Department of Physics, Bose Institute, 93/1 Acharya Prafulla Chandra Road, Kolkata 700009, India}

\begin{abstract}
The detection and management of diseases become quite complicated when pathogens contain asymptomatic phenotypes amongst their ranks, as evident during the recent COVID-19 pandemic. Spreading of diseases has been studied extensively under the paradigm of Susceptible - Infected - Recovered - Deceased (SIRD) dynamics. Various game-theoretic approaches have also addressed disease spread, many of which consider ${\cal S}$, ${\cal I}$, ${\cal R}$, and ${\cal D}$ as strategies rather than as states. Remarkably, most studies from the above approaches do not account for the distinction between the symptomatic or asymptomatic aspect of the disease. It is well-known that precautionary measures like washing hands, wearing masks and social distancing significantly mitigate the spread of many contagious diseases. Herein, we consider the adoption of such precautions as strategies and treat ${\cal S}$, ${\cal I}$, ${\cal R}$, and ${\cal D}$ as states. We also attempt to capture the differences in epidemic spreading arising from symptomatic and asymptomatic diseases on various network topologies. Through extensive computer simulations, we examine that the cost of maintaining precautionary measures as well as the extent of mass testing in a population affects the final fraction of socially responsible individuals. We observe that the lack of mass testing could potentially lead to a pandemic in case of asymptomatic diseases. Network topology also seems to play an important role. We further observe that the final fraction of proactive individuals depends on  the initial fraction of both infected as well as proactive individuals. Additionally, edge density can significantly influence the overall outcome. Our findings are in broad agreement with the lessons learnt from the ongoing COVID-19 pandemic.

\end{abstract}

\maketitle

\section{Introduction}
\label{sec:Intro}
Occasionally, an infectious disease can spread to such an extent, that it significantly affects a vast number of people and indeed the majority of the population. Such a scenario is referred to as an epidemic \cite{daley2001epidemic}. It becomes pandemic in nature when it affects multiple countries, perhaps at the global level. The impact of an epidemic can reach the height of devastation, especially if the pathogen also causes asymptomatic carriers \cite{riggs2007asymptomatic, futai2003role}. Obviously, such asymptomatic  infections present a stark contrast to symptomatic infections --- wherein the symptoms of infections are clearly manifest.  Henceforth, we will refer to the former as class $\cal A$ infections and the latter as class $\cal M$ infections. An infection with asymptomatic phenotypes presents remarkable consequences. It severely hinders the containment of infection and the management of the infected. These unfortunate effects have been in full display during the recent  Covid-19 pandemic \cite{spinelli2020covid} as well as the ``Spanish influenza" \cite{johnson2002updating}. Revolutionary advances in medicine have been achieved between these two catastrophes separated by a century. And yet, little can evidently be done to contain the spread of such pandemics --- when in full rage.

Epidemic spreading has been modelled extensively through SIRD dynamics on networks \cite{kenah2011epidemic,wu2020individual,hu2018individual,wu2018pair}. It has also been studied independently through game-theoretic approaches, many of which involve networks \cite{chang2020game,adiga2020data}. Herein, we adopt  a game-theoretic approach towards epidemic spreading on heterogeneous networks \cite{de2020impact,sharma2019epidemic,chang2020game,castellano2020cumulative,castellano2020cumulative,kenah2011epidemic,li2014analysis}.  
There are not many studies addressing the distinction between class $\cal M$ and class $\cal A$ type infections in existing literature \cite{leung2018infector,choi2020optimal}. One of our prime objectives is to clearly demonstrate the effects arising from differences in asymptomatic and symptomatic phenotype on epidemic modeling.

Furthermore, in existing literature on game theoretic modeling of epidemic spreading on networks $\cal S$, $\cal I$, $\cal R$ and $\cal D$ have been predominantly considered as strategic options \cite{chang2020game}. However, herein we consider these as states rather than strategies. Conversion of an individual's condition from susceptible to infected during epidemic spreading is not always a choice. Therefore, $\cal S$, $\cal I$, $\cal R$ and $\cal D$ are perhaps more aptly considered as states of an individual. On the other hand, individuals may either cooperate or defect in being precautious. Such proactiveness of individuals is more of a choice and could be considered as a strategy rather than a state. This rather important biological aspect has not been accorded sufficient attention in studies on epidemic spreading involving game-theoretic approaches on networks  and SIRD dynamics.

Generally, to prevent the spread of any transmissible disease, mass administration of vaccines  -- if available -- is the most desirable option. Significant research has been conducted to scrutinize the vaccination process and its effect on the spread of an epidemic \cite{hoelscher2008vaccines, sharma2019epidemic}. However, in the absence of vaccines, asymptomatically infected individuals (especially in class $\cal A$ diseases), are likely to be unaware of their own condition.  They can unwittingly act as spreaders or even as super-spreaders if they do not resort to proper precautionary measures \cite{bai2020presumed}.  As is well-known, precautionary measures in the form of frequently washing one's hands, wearing masks, and social distancing are some effective strategies to limit the spread of many contagious diseases.  In more extreme cases, lockdown may be required, which obviously presents significant economic and social costs.  

Susceptible-Infected-Recovered-Deceased or SIRD dynamics specifies that every member of the population would be in one of the following {\em four states}, at any given instant of time, $t$.  Borrowing from the notions of set theory, we can pool individuals in a similar state as belonging to the same set.  Sets representing these states are susceptible, ${\cal S}(t)$, infected, ${\cal I}(t)$, recovered, ${\cal R}(t)$, and deceased, ${\cal D}(t)$ and would also be frequently referred to hereinafter as ${\cal S}$, ${\cal I}$, ${\cal R}$, and ${\cal D}$ respectively \cite{liu2017infectious}. If ${\mathscr E}$ is the universal set denoting the whole population, then obviously ${\mathscr E} = {\cal I} \cup {\cal S} \cup {\cal R} \cup {\cal D}$.

Irrespective of their state, susceptible individuals and infected yet undetected individuals can choose from either of the following two strategies when facing an epidemic. These individuals can either adopt precautionary measures and act responsibly or they can remain careless and spread the disease. Any proactive person would obviously need to pay a certain ``cost" in order to sustain these measures and act responsibly. In addition, mass testing would surely go a long way in identifying infected individuals. Mass testing ratio, $\tau$, denotes the fraction of the entire population, which has been tested for a given disease. Its importance in class $\cal A$ diseases especially can hardly be overemphasised.  A prime motivation of this work is to examine the manner in which precautionary measures and mass testing can affect the spread of an epidemic.

In any infection, the infected individuals definitely possess the capacity to be quite contagious during the incubation period. This is true for both class $\cal A$ and class $\cal M$ diseases. Generally, most diseases are neither purely symptomatic nor purely asymptomatic. However, many diseases are predominantly symptomatic or predominantly asymptomatic. During the course of an epidemic some infected individuals may exhibit symptoms, while many others may not. The manifestation of symptoms may depend on the nature and severity of the disease, the immunity of the infected individuals, and sundry other factors. Herein, we distinguish between class $\cal M$ and class $\cal A$ diseases by using the parameter, $\sigma$, which is the \underline{\bf S}ymptom \underline{\bf M}anifestation \underline{\bf R}atio.  Of all infected individuals,  $\sigma$ is the fraction of infected individuals who clearly demonstrate the symptoms of a given disease. $\sigma=0$ represents the scenario, where all infected individuals are asymptomatic. On the other hand, $\sigma=1$ signifies that all infected individuals clearly exhibit symptoms of a given disease. Generally, $0 < \sigma < 1$. The estimate of $\sigma$ for a given disease can be obtained from empirical data. 

The decisive role of topology is well known in networks in fields as diverse as infrastructure, image processing, optogenetics and phage-bacteria interactions\cite{key2012, PRE_2015, SR_2015, BI_2015, deb2020residue, BI2021}.  The dependence of the outcome of games played on  heterogeneous structures depends non-trivially on the underlying topology \cite{chang2020game, SS_2019, nath2021scalefree, sinha2020topology}. To understand the role of the underlying topology on the dynamics, we consider various population structures in our simulations.  We have predominately studied the dynamics on Barab\'asi-Albert (BA) networks, which is known to emulate many real-world scenarios. In addition, we have also studied the dynamics on Watts–Strogatz small-world (SW) and  Erd\"{o}s-R\'{e}nyi (ER) models. 

\section{Model}
\label{sec:Model}

Our model is quite general and addresses both symptomatic (class $\cal M$) and asymptomatic (class $\cal A$) diseases.  Contagious diseases can spread through several different mechanisms. These could range from being in the mere proximity of an infected individual or coming in contact with contaminated items used or even touched by the individual. It could also spread through the bodily secretions of an infected person, through sexual or other physical contact or through various vectors. 

Proactive individuals will choose precautionary measures of their own accord. Examples of such precautionary measures are the maintenance of social distance, wearing masks, and frequently washing hands. These measures would surely lower the rate of the infection spreading.  However, adopting these measures comes at the cost of a varying degree of restriction in one's everyday life. On the other hand, a careless individual will not choose adequate or proper protection against infection. Such individuals are not merely a danger to many others but even to themselves. Their uncooperative and carefree behaviour is likely to increase the rate of spread of infection. 

Thus, we observe that broadly two strategies are possible  -- cooperation and defection \cite{rowlett2020decisions}. Proactive individuals are cooperators $C$, who act towards preventing the spread of infection by taking proper precaution. However, this comes at a cost, $c$.  If an individual's neighbor happens to be a cooperator, it automatically obtains a benefit $b$. On the other hand, careless individuals can be considered as defectors $D$. They will not spend any cost and therefore, for such defectors,  $c=0$. In spite of their callous behaviour -- these defectors still enjoy benefit, $b$, due to the accommodating behaviour of cooperators in the population. Since this benefit comes at no cost, these defectors effectively act like free-riders in a population.  When two cooperators, $C$, interact -- they will be mutually benefited by each other's proactive actions. However, this benefit has been arrived at by expending a cost $c$. Therefore,  the actual reward enjoyed by these cooperators is $(b-c)$. 

The interaction between two defectors, $D$, will not lead to any benefit or cost, for either of them. Thus, this will result in zero payoffs. For the interaction between a cooperator and a defector;  $D$ will obtain temptation amounting to $b$, while $C$ will derive the sucker's payoff equaling $-c$. True to their nature, defectors would try to avoid paying any cost. Table \ref{tab:payoff} displays the payoff matrix incorporating all the above interactions modeled on the Prisoner's Dilemma.

\begin{table}
\centering
\begin{tabular}{|c|c|c|}
\hline
 & $C$ & $D$  \\
\hline
$C$ & $b-c$, $b-c$ & $-c$,$b$ \\
\hline
$D$  & $b$,$-c$ & $0$,$0$ \\
\hline
\end{tabular}
\caption{Payoff matrix, indicating the payoffs of both the row and the column players. Here, $b=1$ and $0<c<1$. We have considered $c=0$ for ${\cal S}_D$, as they largely do not resort to precautionary measures. Successful vaccination should lead to $c=1$, thereby implying safety from the disease.  $c<1$ would always entail some risk. $b>c$ ensures that the game being played is Prisoner's Dilemma.}
\label{tab:payoff}
\end{table}

Asymptomatic individuals in class $\cal A$ diseases are rather dangerous because they do not display clear symptoms of infection and unwittingly act as spreaders or even super-spreaders. Perhaps the only way to identify such infected individuals is through mass testing. Mass testing allows the identification of both symptomatic and asymptomatic infected individuals -- even the mild ones. The infected individuals who have been detected by mass testing, are kept in quarantine and denoted by $\cal {I_Q}$. Individuals in $\cal {I_Q}$ do not take part in the game anymore. For simplicity and without the loss of generalisation, we presume here that they can not harm members of the susceptible population anymore. They can either recover, which depends on recovery rate, $\rho$, or die depending on the death rate, $\delta$. On the other hand, individuals who are infected but remain undetected, $\cal {I_U}$, cause further infection in the population. $\cal {I_U}$ can be of two types, ${\cal I}_C$ and ${\cal I}_D$. Without medical intervention, $\cal {I_U}$ individuals can also recover at a ``self-recovery rate", $\rho_s$. All the states and the possible strategies associated with each state are listed in Table \ref{Table:state}.

\begin{table}[htbp]
\begin{tabular}{|l|l|}
\hline
\textbf{State} & \textbf{Strategic choice} \\ \hline
Susceptible ($\cal S$) & Susceptible but proactive (${\cal S}_C$)\\
  &  Susceptible but careless(${\cal S}_D$) \\ \hline
Infected (${\cal I_U}$) & Infected but proactive (${\cal I}_C$) \\
(undetected, asymptomatic) & Infected but careless (${\cal I}_D$) \\ \hline
Infected (${\cal I_Q}$) &             No active strategic choice \\
(detected, symptomatic) & \\ \hline
Recovered ($\cal R$) &              No active strategic choice \\ \hline
Deceased ($\cal D$) &              No active strategic choice \\ \hline
\end{tabular}
\caption{Different strategies associated with various states.}
\label{Table:state}
\end{table}

We consider that the rate of infection spread will depend on the strategies of individuals. In a population, $\cal S$  and $\cal {I_U}$ individuals are likely to interact with each other causing the spread of infection. Both $\cal S$  and $\cal {I_U}$ individuals can adopt either cooperation or defection as their strategy. ${\cal S}_C$ and ${\cal I}_C$ denote susceptible cooperators and infected cooperators respectively. ${\cal S}_C$ and ${\cal I}_C$ are both proactive and resort to appropriate precaution while interacting with others. 
On the other hand, ${\cal S}_D$ and ${\cal I}_D$ denote susceptible defectors and infected defectors respectively. Cost, $c$, reflects the extent of precaution undertaken by an individual while the benefit obtained from another proactive individual's  precautionary actions  is reflected in  $b$, as aforementioned. The fraction $c/b$ indicates the extent of precaution undertaken by an individual, where obviously $b>c$. 
We can also define risk as $r=(1-\frac{c}{b})$.  $\cal C$ will resort to precaution at some cost to  lower the risk. However, $c=0$ for $\cal D$, implying $r=1$, i.e., high risk. Let $r_i$ and $r_j$ denote the risk related to two interacting individuals $i$ and $j$. If one of these individuals is infected and other one is susceptible, the probability of the susceptible  individual getting infected due to the interaction is $\mu=(r_i \times r_j)$. This indicates that $\mu$ of ${\cal S}_C$ due to interaction with ${\cal I}_C$ is $\mu_{C,C}=(1-\frac{c}{b})(1-\frac{c}{b}) = (1-\frac{c}{b})^2$. Expectedly this is low because both of them have resorted to adequate precaution. However, the probability that an ${\cal S}_D$ individual gets infected due to an interaction with an ${\cal I}_D$ individual is $\mu_{D,D}=1$. Again, this is expectedly high because none of the individuals have resorted to due precaution. Similarly, $\mu_{C,D}=\mu_{D,C}=(1-\frac{c}{b})$. Let us recall that individuals  in the susceptible and infected state can employ either of the two strategies -- cooperation or defection. 
At any time instant, if ${\cal S}_C$ and ${\cal I}_C$ denote the susceptible and infected cooperators respectively, then ${\cal S}_C \cup {\cal I}_C=C$. Similarly, if  ${\cal S}_D$ and ${\cal I}_D$ denote the susceptible and infected defectors, then ${\cal S}_D \cup {\cal I}_D=D$. If ${\cal I}_C$ and ${\cal I}_D$ denote undetected individuals, then ${\cal I}_C \cup {\cal I}_D=\cal {I_U}$. Obviously, ${\cal S} = {\cal S}_C \cup {\cal S}_D$ and ${\cal I} = {\cal I_U} \cup {\cal I_Q}$.

We have already discussed in detail about the importance of SMR, $\sigma$, in Section~\ref{sec:Intro}.  For class $\cal M$ diseases, we have considered $\sigma=0.9$, which is rather high. On the other hand, $\sigma=0.02$, has been considered for class $\cal A$ diseases \cite{chen2020infectious}.

\section{Algorithm}
\label{sec:Algorithm}
In our simulations, a heterogeneous population structure  has been considered in the form of a Barab\'asi-Albert (BA) network, with average degree $\langle k \rangle$.  It is well-known that BA networks possess a power-law degree distribution. Simultaneously, we have also considered small-world and random networks in the form of Watts–Strogatz small-world (SW) and  Erd\"{o}s-R\'{e}nyi (ER) models. Each time step during both transient time and counting time sequentially incorporates the events of: (a) payoff determination, (b) strategy upgradation, and (c) state upgradation. Initially every individual in the population is susceptible. At the start of an epidemic session, some of these individuals get  infected randomly. Thus now, we have two states -- susceptible and infected. The population has been divided into cooperators, $C$, and defectors, $D$. If ${\cal I}_i$ represents the set of initially infected individuals, ${\cal I}_i$ remains undetected i.e., ${\cal I_U}={\cal I}_i$ and ${\cal I_Q}=\emptyset$. Thus, on the basis of both strategy and state, there exist four types of individuals, namely, ${\cal S}_C$, ${\cal S}_D$, ${\cal I}_C$, and ${\cal I}_D$. Initially, prisoner's dilemma is played between $C$ and $D$, irrespective of the state they belong to. They will accumulate their payoffs by interacting with each other. It is rather superfluous to consider any strategy for recovered and dead individuals and they do not take part in the game. After determination of payoffs, all cooperators and defectors randomly select a neighbor for strategy upgradation. Following Fermi's rule of strategy upgradation -- an individual, $i$, will adopt the strategy of a randomly chosen neighbor, $j$, with a probability $\frac{1}{1+exp[-(\Pi_j-\Pi_i)]}$. $\Pi_j$ and $\Pi_i$ denote the total payoffs accumulated by $j$ and $i$ respectively \cite{liu2017fixation,santos2005scale}. Strong selection has been considered here.

After strategy upgradation, state upgradation takes place. In the previous step, $\cal S$  and $\cal {I_U}$ individuals decide to be either $C$ or $D$. But in this step, infection will spread in the population. It has been considered that the infection can spread through contact. Thus, susceptible neighbors of an infected individual are vulnerable. But depending on the strategy of both the interacting individuals i.e., $C$ or $D$, the probability of spreading of infection will be different. 

While upgrading its state, $\cal S$ will randomly interact with one of its neighbors. If that randomly chosen neighbor is $\cal {I_U}$, $\cal S$ may get infected with a probability, $\mu$. The newly infected individuals may or may not manifest the symptoms. This would depend on the symptom manifestation rate of the infection, $\sigma$. The symptomatic individuals would be quarantined and they would not interact with the rest of the population anymore. As mentioned earlier, these individuals are identified as $\cal {I_Q}$. The asymptomatic individuals will remain in population and would cause further infection.

All individuals would hence undergo mass testing with probability, $\tau$.  Depending on $\tau$, some infected individuals can also be identified as $\cal {I_Q}$. The remaining $\cal {I_U}$ would remain in the population. Since they are unaware of their own situation they will unwittingly infect others. They may undergo self-recovery depending on $\rho_s$ or die without getting any medical help, depending on the value of $\delta$. On the other hand, $\cal {I_Q}$ individuals may either recover, depending on the value of $\rho$, or die, depending on the value of $\delta$. Here for simplicity, the value of $\delta$ of $\cal {I_Q}$ and $\cal {I_U}$ has been considered to be the same. In our simulations, reinfection of recovered individuals has not been considered. 

In all simulations, a transient time of $10^4$ generations is considered. After this transient time, if ${\cal N}_{{\cal S}_C}$ be the number of susceptible cooperators, $f_C = {\cal N}_{{\cal S}_C}/{\cal N}$, is calculated over a counting time of $10^3$ generations. Every network has ${\cal N}$ nodes and the average degree is denoted by $\langle k \rangle$. The overall simulations have been performed over $E_{\cal N}$ networks. In all simulations, a definite fraction of nodes is initially assigned to be cooperators, at random. If $f_{C_i}$ denotes the initial fraction of cooperators in the population, then $f_{D_i}=(1-f_{C_i})$ denotes the initial fraction of defectors. Apart from strategies, a definite number of nodes, i.e. individuals, are also randomly chosen to get infected. We denote the initial number of infected and susceptible individuals by ${\cal N}_{{\cal I}_i}$ and ${\cal N}_{{\cal S}_i}$ respectively. Obviously, $({\cal N}_{{\cal S}_i} +{\cal N}_{{\cal I}_i})={\cal N}$ denotes the total number of individuals in the population. The fraction of ${\cal I}_i$ and ${\cal S}_i$ is denoted by $f_{{\cal I}_i}={\cal N}_{{\cal I}_i}/{\cal N}$ and $f_{{\cal S}_i}={\cal N}_{{\cal S}_i}/{\cal N}$ respectively. Here, $f_{{\cal I}_i}=(1-f_{{\cal S}_i})$. At the very onset of the epidemic, all initially infected individuals would be undetected. Hence, initially ${\cal N}_{{\cal I}_{{\cal U}}}={\cal N}_{{\cal I}_i}$ and ${\cal N}_{{\cal I}_{{\cal Q}}}=0$. Here, ${\cal N}_{{\cal I}_{{\cal U}}}$ and ${\cal N}_{{\cal I}_{{\cal Q}}}$ denote the number of undetected and detected individuals respectively.

\section{Results}
\label{sec:Results}

\begin{figure}[h]	
	\includegraphics[width=\columnwidth, height=5cm]{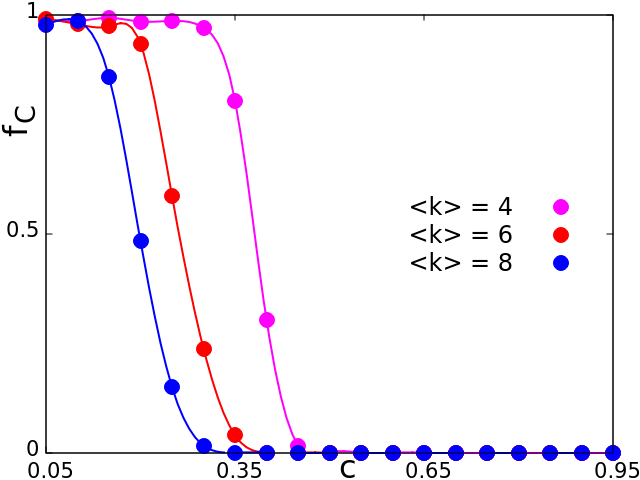}
	\caption{$f_C = {{\cal S}_C}/{\cal N}$, versus cost, $c$. The fraction of population, which was initially infected is $f_{{\cal I}_i}=0$. Here, the initial fraction of cooperators, $f_{C_i}=0.5$. Results are for a network of ${\cal N}=1024$ nodes, $\langle k \rangle = $ $4$, $6$, $8$ and $E_{\cal N}=580$ networks. Evidently, in the absence of infection, all individuals are susceptible, i.e., either ${\cal S}_C$ or ${\cal S}_D$. We observe that cooperation is maintained at lower values of $c$, which is compromised with increasing $c$. Cooperation seems to be maintained relatively better for graphs with lower $\langle k \rangle$. The standard error is smaller than the size of the data points.} 
	\label{fig:only_game} 
\end{figure}

Prisoner's Dilemma game  is  played initially in the absence of any infection, i.e., between ${\cal S}_C$ and ${\cal S}_D$ only, which implies $f_{{\cal S}_i}=1$ and $f_{{\cal I}_i}=0$. In that case, $f_C = {\cal N}_{{\cal S}_C}/{\cal N}$, depends only on $c$. ${\cal S}_C$ dominates at lower values of $c$ but decreases at higher values of $c$, as shown in Fig \ref{fig:only_game}. This implies that if $c$ increases, individuals would be quite unwilling to adopt precautionary measures and would try to defect.

\subsection{Asymptomatic infections: class $\cal A$}
\begin{figure}[h]	
	\includegraphics[width=\columnwidth, height=5cm]{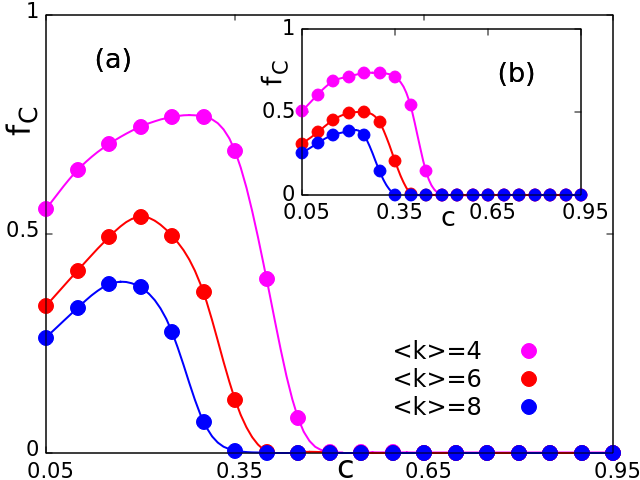}
	\caption{$f_C$ versus $c$  for a predominantly asymptomatic infection with $\sigma=0.02$. Results are for (a) ${\cal N}=1024$ nodes and (b) ${\cal N}=2048$ nodes. Other parameters are, $f_{C_i}=0.5$, $f_{{\cal I}_i}=0.02$, $\rho=0.5$, $\rho_s=0.2$, $\tau = 0.3$, $\delta=0.05$,  $\langle k \rangle = $ $4$, $6$, $8$ and $E_{\cal N}=580$ networks. Cooperation is witnessed to some extent in a limited range of $c$. In asymptomatic (class $\cal A$) diseases, lower $c$ indicates a higher risk of getting infected. On the other hand, a higher cost is naturally unaffordable to most individuals. The standard error is smaller than the size of the data points.}
	\label{fig:asym} 
\end{figure}

 First, we have considered class $\cal A$, where the symptom manifestation ratio, $\sigma$, is very low. There are various diseases where  manifestation of symptoms can not often be observed clearly, such as infections associated with Cytomegalovirus, Rhinoviruses, Salmonella or Ebola \cite{kemper1978effects, de2017new, potasman2017asymptomatic} infections. In the Zika virus infection, symptoms of clinical illness were absent in more than 80\% of infected patients. These individuals played a crucial role in the initial transmission of the Zika virus \cite{moghadas2017asymptomatic}. Similarly, the asymptomatic pine trees, harbouring {\it B. xylophilus} nematodes play a vital role in the spreading of  pine wilt epidemic \cite{futai2003role}. The ongoing pandemic due to COVID‐19 virus is also associated with a high number of asymptomatic carriers \cite{nishiura2020estimation}.  
 Herein, we have considered that initially $2\%$ individuals have been infected. Fig. \ref{fig:asym} is for the asymptomatic case. It can be observed that $f_C$ decreases with the increase of $c$. Hence, it can be concluded that individuals in the population will only cooperate for low value of $c$. But due to the presence of infection, low $c$ implies higher risk. 

Hence, alongside defectors, these cooperators are also likely to get infected at lower values of $c$. However, if we increase $c$, better protection ensures higher $f_C$. For a $BA$ network with $\langle k \rangle=4$, we observe that $f_C$ is higher when $c=[0.25,0.35]$. Though a higher cost implies lower risk, it also inhibits the lifestyle of ${\cal S}_C$ individuals. Naturally, cooperation would be hindered at higher values of $c$. Hence, with increasing costs, ${\cal S}_C$  individuals tend to becomes careless, implying that ${\cal S}_D$ individuals would be likely to dominate. Also ${\cal S}_D$ individuals would get infected at a higher rate due to the absence of proper precaution on their part. In the case of high $\langle k \rangle$ graphs, as shown in Fig. \ref{fig:asym}, it has been observed the peak shifts towards the lower value of $c$. Generally, the maintenance of cooperation is lower when $\langle k \rangle$ of the graph is high. 

\begin{figure}[htbp]
	\centering
	\includegraphics[width=\columnwidth, height=5cm]{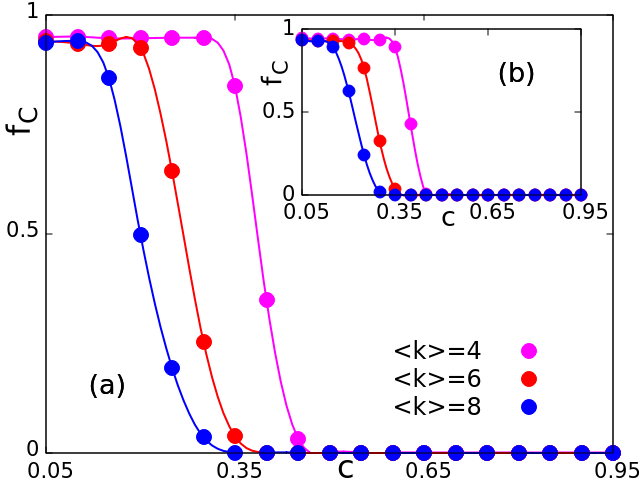}
	\caption{$f_C$ versus $c$,  for a predominantly symptomatic disease with $\sigma=0.9$. Results are for (a) ${\cal N}=1024$ nodes and (b) ${\cal N}=2048$ nodes. Other parameters are, $f_{C_i}=0.5$, $f_{{\cal I}_i}=0.02$, $\rho=0.5$, $\rho_s=0.2$, $\tau = 0.3$, $\delta=0.05$,  $\langle k \rangle = $ $4$, $6$, $8$ and $E_{\cal N}=580$ networks. Increasing costs lead to decreasing cooperation. The standard error is smaller than the size of the data points.} 
	\label{fig:symp} 
\end{figure}

\subsection{Symptomatic infections: class $\cal M$}
Class $\cal M$ diseases obviously have a high symptom manifestation ratio associated with them. Infections like {\it Vibrio cholerae} and Poxviruses are examples of such symptomatic diseases \cite{lewis2004zoonotic}. This scenario has been examined in Fig. \ref{fig:symp}. Here too, $f_C$ is observed to decrease with an increase of $c$ at different values of $\langle k \rangle$. 

\begin{figure}[htbp]
	\centering
	\includegraphics[width=\columnwidth, height=7cm]{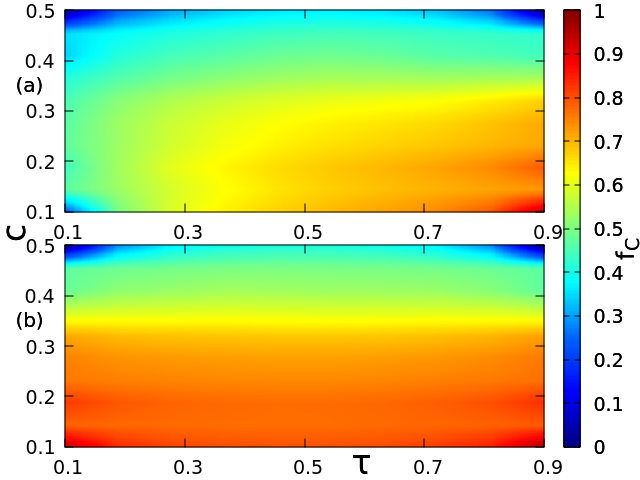}
	\caption{$f_C$ versus $\tau$ at various values of $c$ as quantified for: (a) $\sigma = 0.02$ (class $\cal A$), and, (b) $\sigma = 0.9$ (class $\cal M$). Red represents maintenance of cooperation and blue it's absence. For class $\cal A$ diseases, cooperation depends on both $\tau$ and $c$. Higher value of $\tau$ can lead to more  cooperation. Cooperation will not be maintained well enough at lower values of $\tau$. However, for class $\cal M$ diseases, maintenance of cooperation depends more on $c$ rather than $\tau$. This indicates that mass testing plays an important role for the survival of cooperators in the case of class $\cal A$ diseases. Results are for $f_{{\cal C}_i}=0.5$, $\delta=0.05$,$f_{{\cal I}_i}=0.02$, $\rho=0.5$, $\rho_s=0.2$, ${\cal N}=1024$, $\langle k \rangle=4$, and $E_{\cal N}=100$.}
	\label{fig:heat} 
\end{figure}

\subsection{Comparing symptomatic and asymptomatic infection} 
\begin{figure}[htbp]	
	\includegraphics[width=\columnwidth, height=5cm]{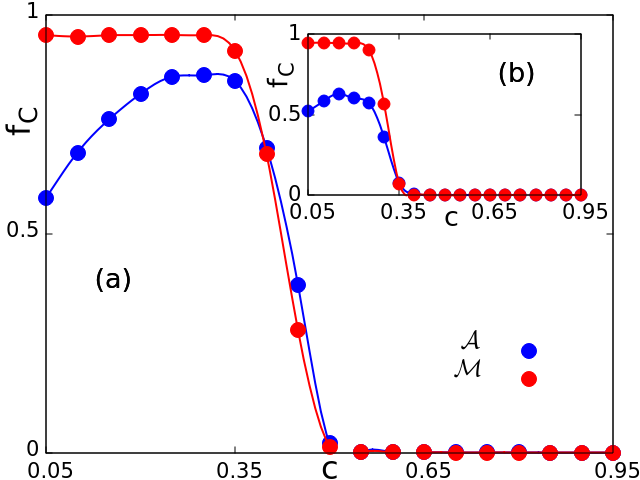}
	\caption{$f_C$ versus $c$  for predominantly symptomatic (class ${\cal M}$, $\sigma=0.9$)  and predominantly asymptomatic (class ${\cal A}$, $\sigma=0.02$) diseases. We study the effect of the initial fraction of cooperators in the population, $f_{C_i}$, on the final fraction, $f_C$, for: (a) $f_{C_i}=0.7$ and (b) $f_{C_i}=0.3$ in both class $\cal A$ and class $\cal M$. Results are for $f_{{\cal I}_i}=0.02$, $\rho=0.5$, $\rho_s=0.2$, $\tau = 0.3$, $\delta=0.05$, ${\cal N}=1024$ nodes, $\langle k \rangle = $ $4$ and $E_{\cal N}=500$ networks. The standard error is smaller than the size of the data points.}
	\label{fig:percentage} 
\end{figure}

We study the fraction of cooperators, $f_C$, in both symptomatic and asymptomatic infections at $f_{C_i}=0.5$ in Fig. \ref{fig:heat} for a range of values of cost, $c$. In order to find how $f_C$ varies with the initial fraction of cooperators we measure $f_C$ for $f_{C_i}=0.7$ in Fig. \ref{fig:percentage}(a) and $f_{C_i}=0.3$ in Fig. \ref{fig:percentage}(b). We observe that the value of the initial fraction of cooperators does affect the final fraction of proactive individuals to an extent.

From Fig. \ref{fig:heat} it is also evident that both mass testing and cost due to precautionary measures have a significant impact on $f_C$. At higher values of $c$, cooperation will not be maintained in the population. But at lower values of $c$, the maintenance of cooperation would depend on the value of $\tau$. In class $\cal A$ diseases, it is difficult to maintain cooperation at lower values of $\tau$. This can be intuitively inferred from the fact that it is rather difficult to identify infected individuals without mass testing. Therefore, higher values of $\tau$ can lead to relatively higher cooperation. On the other hand, $\tau$ does not really have a significant impact in class $\cal M$ diseases.

Therefore, it is worth noting that when  the mass testing ratio, $\tau$, is low, an asymptomatic disease is likely to infect susceptible individuals more. Naturally, if the infected individuals are not identified early on,  then every asymptomatic disease possess the potential to generate a pandemic. On the other hand, individuals belonging to $I_U$ in class $\cal M$ diseases, can be identified more easily via clearly manifest symptoms. 

\begin{figure}[htbp]
	\centering
	\includegraphics[width=\columnwidth, height=7cm]{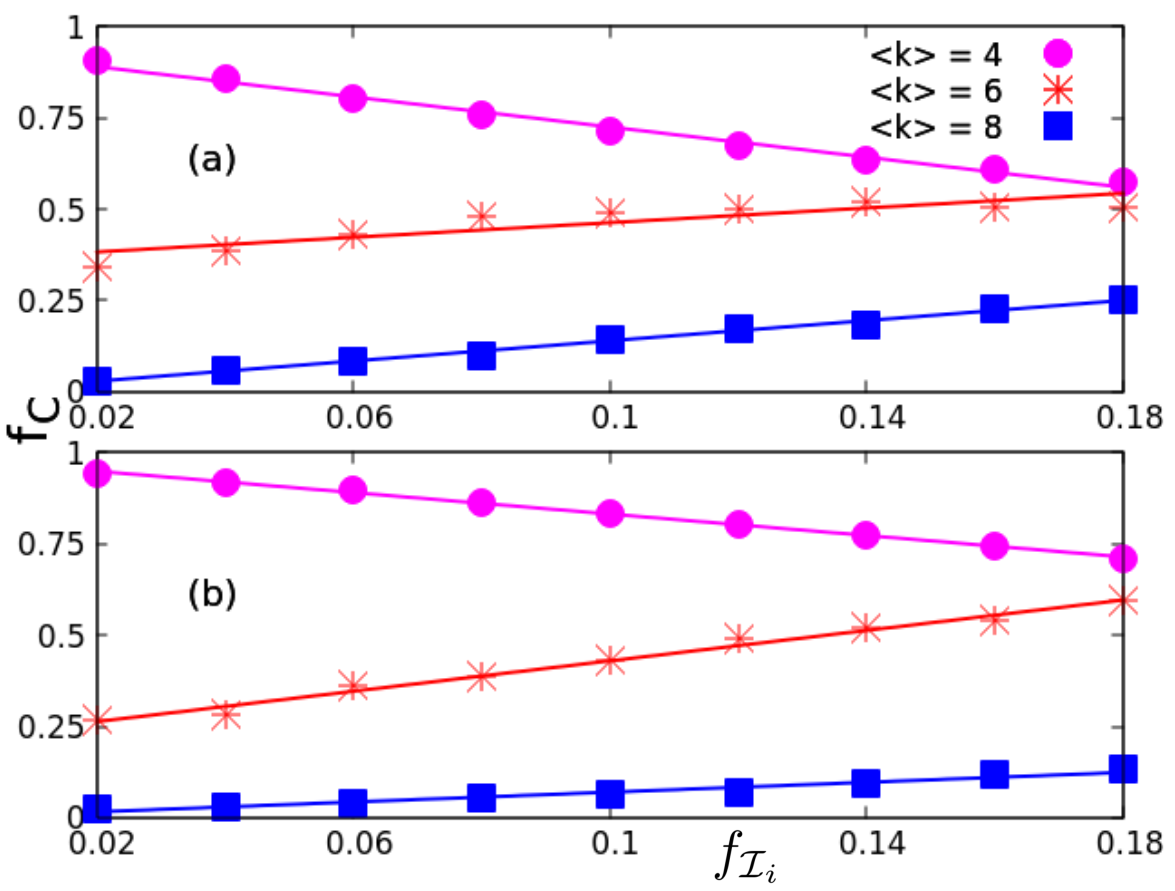}
	\caption{$f_C$ versus $f_{{\cal I}_i}$, as quantified for: (a) $\sigma = 0.02$ (class $\cal A$), and, (b) $\sigma = 0.9$ (class $\cal M$).  Results are for $c=0.3$, $f_{C_i}=0.5$, $\rho=0.5$, $\rho_s=0.2$, $\tau = 0.3$, $\delta=0.05$, ${\cal N}=1024$, $\langle k \rangle =$ $4$, $6$, $8$ and $E_{\cal N}=500$.  The fraction of proactive individuals actually decreases with the increase of ${\cal I}_i$ at lower $\langle k \rangle$. However, cooperation increases slightly with the increase in ${\cal I}_i$ at higher $\langle k \rangle$. The standard error is smaller than the size of the data points.}
	\label{fig:ii} 
\end{figure}

The initial fraction of infected individuals, $f_{{\cal I}_i}$, reflects the scenario during the initial outbreak of an epidemic. We explore the behaviour of $f_C$ with respect to $f_{{\cal I}_i}$ \cite{sahneh2012existence}.  A fraction of the population, $f_{{\cal I}_i}$,  is infected right at the onset of the epidemic,  before  consciousness has taken root in the general population. Consciousness is expected to increase subsequently and the general populace would become aware and responsible. Fig. \ref{fig:ii} implies that  the final fraction of proactive individuals would decrease in the population at lower $\langle k \rangle$. However, at higher $\langle k \rangle$, the scenario is not quite similar. In this regime,  the final fraction of responsible people would increase slightly in the population. Notably, the broad behaviour is the same for both class $\cal M$ and class $\cal A$ epidemics. 

\begin{figure}[h]	
	\includegraphics[width=\columnwidth, height=5cm]{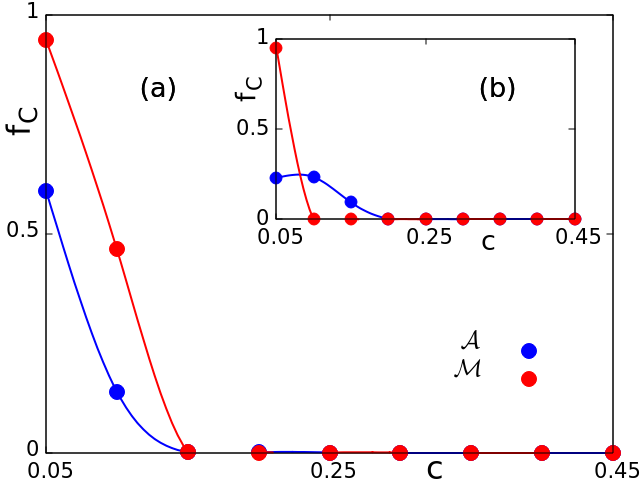}
	\caption{$f_C$ versus $c$  for predominantly symptomatic (class ${\cal M}$, $\sigma=0.9$)  and predominantly asymptomatic (class ${\cal A}$, $\sigma=0.02$) diseases. Results are for (a) Watts-Strogatz network with ${\cal N}=1024$ nodes, $\langle k \rangle=10$, edge rewiring probability $\beta=0.001$, and, (b) Erd\"{o}s-R\'{e}nyi network with $N=512$ nodes, $p=0.005$. Here, $f_{C_i}=0.5$, $f_{{\cal I}_i}=0.02$, $\rho=0.5$, $\rho_s=0.2$, $\tau = 0.3$, $\delta=0.05$ and $E_{\cal N}=500$ networks. The standard error is smaller than the size of the data points.}
	\label{fig:sw_erdos} 
\end{figure}

\subsection{Role of network topology}
Thus far, we have extensively studied the dynamics on BA networks. We would further like to ascertain the role of network topology on the dynamics of epidemic spreading. Towards this end, we have studied the dynamics on  Watts-Strogatz small-world networks (SW) and Erd\"{o}s-R\'{e}nyi (ER) networks as demonstrated in Figs. \ref{fig:sw_erdos}(a) and \ref{fig:sw_erdos}(b) respectively. In comparison with the earlier results on BA networks, we observe the final fraction of proactive individuals to vanish at far lower values of the cost in both ER and SW networks for both class ${\cal M}$ and class ${\cal M}$. This effect is most prominent for class ${\cal M}$ diseases in ER networks. Therefore, population structures on SW and more so on ER networks seem vulnerable to pandemics. 

\section{Discussion}
Accurate models of epidemic spreading are essential \cite{liang2020mathematical,mcbryde2020role,currie2020simulation}, as has been clearly underscored during the ongoing COVID-19 pandemic, which has raged for more than a year now and caused enormous setback to life, economy and society. Prevalence of asymptomatic carriers complicates the detection and management of any disease to a significant extent \cite{yu2020covid,recalcati2020cutaneous,lai2020asymptomatic}. Indeed, the crucial role of asymptomatic carriers has been clearly highlighted in the ongoing COVID-19 pandemic.

Models based on SIRD dynamics or game-theoretic approaches are not be able to capture the complete picture as they usually do not consider the important aspect of asymptomatic carriers \cite{iwamura2018realistic, chang2020game}. Our model explores the possible effect of various factors in epidemic spreading on various network topologies. We find that the absence of vaccines makes  individual and social proactiveness  as important factors, which can inhibit disease spreading. 

Also, the fraction of the population which has undergone testing is an essential factor in diseases involving asymptomatic phenotypes. In the absence of high enough testing --  an asymptomatic disease bears the potential to get converted into a pandemic. We also observe that the fraction of proactive individuals depends upon  the initial fraction of both infected as well as proactive individuals. Additionally, edge density can significantly  influence the dynamics of epidemic spreading. Furthermore, population structures on small-world networks and  more so on Erd\"{o}s-R\'{e}nyi networks seem vulnerable to pandemics.

Our findings are in broad agreement with the lessons learnt from the ongoing COVID-19 pandemic. Adoption of precaution and employment of intelligent strategies to conduct mass testing \cite{bbcIndia, zilinskas_pooled_2021} has indeed proven to be of immeasurable value \cite{TheGuardian, BBC_health,pnas2021}  towards effective management of the pandemic.

\section{Acknowledgements}
DN acknowledges his fellowship from the Council of Scientific  and Industrial Research (CSIR), India. SR thanks Tarik Hadzibeganovic of University of Graz, Austria for helpful discussions. We are grateful to the three anonymous reviewers of the manuscripts for their critical comments.\\

DN and SS contributed equally to this work.  

\bibliography{Epidemic_Clean}
\bibliographystyle{spbasic}

\end{document}